\documentclass{iopart}
\usepackage{euscript,iopams,amssymb,amsfonts,graphicx,bm}
\usepackage{pgfplots,setstack}

\usepackage{float}
\usepackage{epsfig}
\usepackage{esint}
\usepackage{chemfig}
\usepackage{cite}

\usepackage{bm,braket}

\renewcommand{\theequation}{\arabic{section}.\arabic{equation}}

\newcommand{\calU}{{\mathcal U}}

\newcommand{\calD}{{\mathcal D}}
\newcommand{\calP}{{\mathcal P}}

\newcommand{\calF}{{\mathcal F}}
\newcommand{\calJ}{{\mathcal J}}
\newcommand{\R}{{\mathbb R}}
\renewcommand{\L}{{\mathbb L}}
\newcommand{\X}{\mathbf{X}}

\renewcommand{\P}{\mathbb{P}}

\newcommand{\x}{\mathbf{x}}
\newcommand{\y}{\mathbf{y}}
\renewcommand{\e}{{\mathrm e}}
\newcommand{\E}{{\mathbb E}}

\newcommand{\n}{\mathbf n}
\newcommand{\calT}{{\mathcal T}}
\newcommand{\calS}{{\mathcal S}}

\newcommand{\calN}{{\mathcal N}}

\newcommand{\calK}{{\mathcal K}}

\renewcommand{\P}{\mathbb P}

\newcommand{\well}{h}
\newcommand{\fT}{{\mathfrak T}}
\renewcommand{\S}{\widetilde{S}}

\eqnobysec

\begin{document}

\title[Run-and-tumble motion with diffusion]{Run-and-tumble particle with diffusion: boundary local times and the zero-diffusion limit}

\author{Paul C. Bressloff}
\address{Department of Mathematics, Imperial College London, London SW7 2AZ, UK}
\ead{p.bressloff@imperial.ac.uk}

\begin{abstract} Incorporating boundary conditions into stochastic models of passive or active particle motion is usually implemented at the level of the associated forward or backward Kolmogorov equation, whose solution determines the probability distribution of sample paths. In order to write down the corresponding stochastic differential equation (SDE) that generates the individual sample paths, it is necessary to introduce 
 a Brownian functional that keeps track of the boundary-particle contact time. We previously constructed the SDE for a (non-diffusing) run-and-tumble particle (RTP) on the half-line with either a reflecting or sticky wall at $x=0$. In this paper we extend the theory to include the effects of diffusion. One of the non-trivial consequences of combining drift-diffusion with tumbling is that the zero diffusion limit $D\rightarrow 0$ is singular in the sense that the number of boundary conditions is doubled when $D>0$. We use stochastic calculus to derive the forward Kolmogorov equation for two distinct boundary conditions that reduce, respectively, to non-sticky and sticky boundary conditions in the zero-diffusion limit. In the latter case, it is necessary to include a boundary layer in a neighbourhood of the wall and use singular perturbation theory. 
We also treat the wall as partially absorbing by assuming that the particle is absorbed when the amount of boundary-particle contact time (discrete or continuous local time) exceeds a quenched random threshold. 
 Finally, we analyse the survival probability and corresponding FPT density for absorption at a non-sticky wall
by deriving the corresponding backward Kolmogorov equation.

 \end{abstract}

\maketitle

  \section{Introduction} 
 
 Incorporating boundary conditions into stochastic models of passive or active particle motion is usually implemented at the level of the associated forward Kolmogorov equation (or its backward version), whose solution determines the probability distribution of sample paths. For example, consider pure Brownian motion in a bounded domain $\calU\subset \R^d$ with a partially absorbing boundary $\partial \calU$. The diffusion equation for the probability density $p(\x,t)$ takes the form $\partial_tp(\x,t)=D{\bm \nabla^2}p(\x,t)$, $\x \in \calU$, supplemented by the Robin boundary condition $D{\bm \nabla}p(\y,t)\cdot \n(\y) =-\kappa_0p(\y,t)$, $\y \in \partial \calU$, where $\n(y)$ is the outward unit normal at a point on the surface, $D$ is the diffusivity and $\kappa_0$ is a constant reactivity. The totally reflecting and totally absorbing boundary conditions are obtained by setting $\kappa_0=0$ and $\kappa_0\rightarrow \infty$, respectively. In order to write down the corresponding stochastic differential equation (SDE) that generates the individual sample paths, it is necessary to introduce 
 a Brownian functional known as the boundary local time $\ell(t)$ \cite{McKean75}. This is a continuous, positive, non-decreasing stochastic process that essentially measures the amount of contact time between a particle and the boundary. Given the position $X(t)\in \calU$ of the particle at time $t$, the local time is defined as
 \begin{equation*}
\fl \ell(t)=\lim_{\epsilon\rightarrow 0} \frac{D}{\epsilon} \int_0^t\Theta(\epsilon-\mbox{dist}(\X(\tau),\partial \calU)d\tau=\int_0^t \left \{ \int_{\partial \calU}\delta(\X(\tau)-\y)d\y \right \}d\tau,
\end{equation*} 
where $\Theta(x)$ is the Heaviside function and $\delta(\x)$ is the multidimensional Dirac delta function. The SDE for totally reflected Brownian motion is given by the so-called Skorokhod SDE \cite{Grebenkov06} 
$dX(t)=\sqrt{2D}dW(t)+d\ell(t)$,
where $W(t)$ is a Wiener process with $\langle W(t)\rangle =0$ and $\langle W(t)W(t')\rangle = \min\{t,t'\}$, Sample paths of the classical Robin boundary value problem are then realised by adding the condition that the particle is absorbed when its local time exceeds an exponentially distributed quenched random threshold \cite{Grebenkov06}. Allowing for non-exponentially distributed thresholds is the basis of encounter-based models of diffusion-mediated absorption \cite{Grebenkov20,Grebenkov22,Bressloff22,Bressloff22a,Grebenkov24}.

\begin{figure}[t!]
  \centering
   \includegraphics[width=12cm]{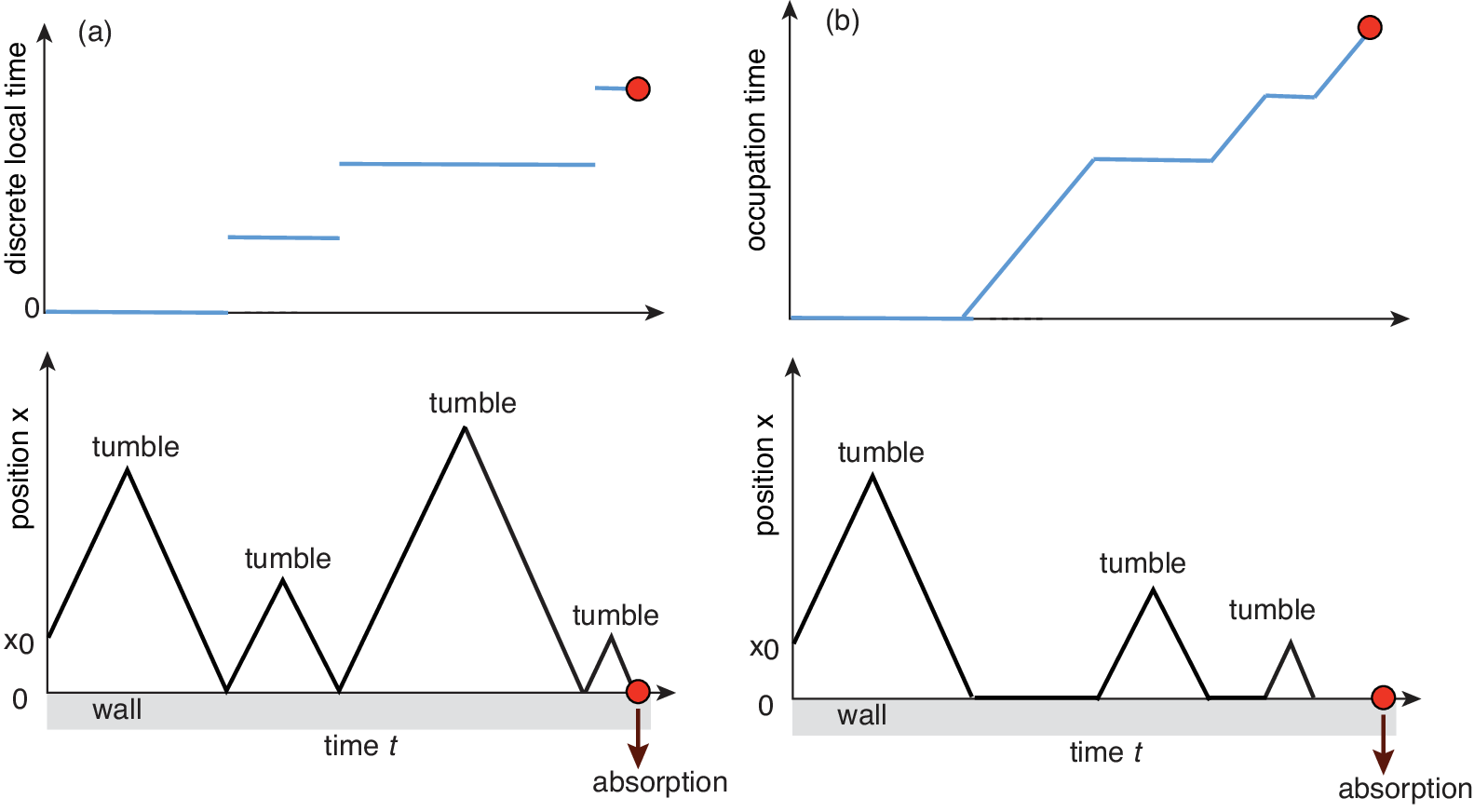}
  \caption{RTP without diffusion confined to the half-line with a partially absorbing wall at $x=0$. (a) Sample trajectory in the case of a {\em non-sticky wall}. The corresponding jumps in the discrete local time are also shown. The local time counts the number of wall collisions over the time interval $[0,t]$. Absorption occurs when the local time crosses a quenched discrete random threshold. (b) Sample trajectory in the case of a {\em sticky wall} at $x=0$. The corresponding piecewise continuous occupation time (time spent attached to the wall) is also shown. The particle is absorbed when the occupation time crosses a quenched continuous random threshold $h$.}
  \label{fig1}
  \end{figure}

We have recently developed an analogous theory for the sample paths of a run-and-tumble particle (RTP) confined to the half-line $[0,\infty)$ with a partially absorbing wall at $x=0$ \cite{Bressloff22b,Bressloff25a}. Let $X(t)$ and $v\sigma(t)$ denote the position and velocity of the particle at time $t$, where $\sigma(t)\in \{-1,+1\}$ is a dichotomous noise process. Away from the wall we have $dX(t)=v\sigma(t)dt$ with $\sigma(t)\rightarrow -\sigma(t)$ at a random sequence of times generated from a homogeneous Poisson process with rate $\alpha$. In the case of a totally reflecting (non-sticky) wall, the particle instantaneously reverses direction from $\sigma=-1$ to $\sigma=+1$ whenever it hits the wall, see Fig. \ref{fig1}(a). In contrast to a Brownian particle, the local time is now a discrete process that counts the number of wall collisions over a given time interval $[0,t]$. The analog of the Robin boundary condition is obtained by assuming that absorption occurs when the discrete local time exceeds a geometrically distributed quenched discrete random threshold. Using the stochastic calculus of jump processes, we showed that the corresponding distribution of sample paths is given by the solution of the forward Kolmogorov equation \cite{Bressloff25a}
\numparts
\begin{eqnarray}
\label{KA0}
&\frac{\partial p_1}{\partial t}=-v \frac{\partial p_1} {\partial x}+\alpha [p_{-1}-p_1],\quad x >0,\\
\label{KB0}&\frac{\partial p_{-1}}{\partial t}=v \frac{\partial p_{-1}} {\partial x}-\alpha [p_{-1}-p_1],\quad x >0,\\
\label{KC0}&v[p_1(0,t)-p_{-1}(0,t)]=-\kappa_0p_1(0,t),
\end{eqnarray}
\endnumparts
where 
\[p_k(x,t)dx=\P[x<X(t)<x+dx,\, \sigma(t)=k].\]
As in the case of diffusion processes, one advantage of working at the level of sample paths and local times is that one can develop a more general encounter-based formulation of RTP absorption by taking the distribution of quenched local time thresholds to be non-geometrical. It is also possible to extend the theory to a partially absorbing sticky wall \cite{Bressloff23,Bressloff25b}. If the wall is sticky then the particle remains attached to the wall until it reverses its velocity state at a rate that might differ from the tumbling rate in the bulk, see Fig. \ref{fig1}(b). Absorption now depends on the amount of time the particle is attached to the wall (continuous occupation time). 

The forward or backward Kolmogorov equation is the usual starting point for solving the first passage time (FPT) problem at a totally absorbing or partially absorbing wall in 1D \cite{Angelani15,Angelani17,Dhar19,Bressloff22b,Angelani23,Gueneau24,Gueneau25}. Such studies have also been extended to include the effects of stochastic resetting \cite{Evans18,Bressloff25a}, spatial inhomogeneities \cite{Singh20}, higher spatial dimensions \cite{Mori20,Santra20,Santra20a,Doussal22,Pal23}, and diffusion \cite{Malakar18}. In the last example, the corresponding SDE away from a boundary wall becomes $dX(t)=v\sigma(t)dt+\sqrt{2D}dW(t)$. One of the non-trivial consequences of combining drift-diffusion with tumbling is that the zero diffusion limit $D\rightarrow 0$ is singular in the sense that the number of boundary conditions is doubled when $D>0$. Moreover, Fig. \ref{fig1} suggests that there are at least two different choices of boundary conditions when $D>0$, whose zero diffusion limits recover the non-sticky and sticky boundary conditions, respectively.
Deriving these boundary conditions from first principles and understanding the singular limit are the main goals of the current paper. 

The structure of the paper is as follows. In section 2, we formulate the SDE of an RTP with diffusion on the half-line and a hard wall at $x=0$. This requires introducing a pair of weighted local times $\ell_{\pm}(t)$ that depend on whether the particle is in the $\sigma(t)=1$ or $\sigma(t)=-1$ velocity state. We distinguish between two different models of reflected motion. In model A, the particle reverses its velocity state each time it collides with the wall in the $\sigma=-1$ state. The corresponding local time $\ell_-(t)$ is then discrete and is related to the number $L(t)$ of collisions in the $(-)$ state. Model B excludes reversals in the velocity state at the wall so that $\ell_-(t)$ is continuous. (The corresponding local time $\ell_+(t)$ is continuous in both cases.) We then extend the SDE to the case of a partially absorbing wall by formulating the first passage time problem in terms of a quenched local time threshold $h$. In model A we assume that the particle is absorbed at the stopping time $\calT=\inf\{t>0,\, L(t)>h\}$ where $h$ is a discrete random variable. On the other hand, in model B the stopping time is taken to be $\calT=\inf\{t>0,\, \ell_+(t)+\ell_-(t)>h\}$ with $h$ a continuous random variable. 

In section 3, we use the stochastic calculus of jump-diffusion processes to derive the forward Kolmogorov equation for models A and B. In particular, we show that the boundary conditions at $x=0$ depend on the particular model of reflected motion at the wall prior to absorption and these have distinct zero-diffusion limits. In the case of model A, the  boundary conditions reduce to those previously derived for an RTP without diffusion and a non-sticky wall \cite{Bressloff25a}. On the other hand, the boundary conditions of model B are singular in the limit $D\rightarrow 0$. The latter can be handled by introducing a boundary layer in a neighbourhood of the wall, which is equivalent to treating the wall as a sticky boundary. Finally, in section 4 we analyse the survival probability and corresponding FPT density for the half-line.
The survival probability is the solution of the corresponding backward Kolmogorov equation, which we derive from first principles. We show that model A recovers the survival probability of a non-sticky absorbing wall in the limit $D\rightarrow 0$.

\setcounter{equation}{0}

\section{RTP with diffusion on the half-line }

Consider a single RTP moving on the half-line $\R^+=[0,\infty)$ with a wall at $x=0$. Let $X(t)\in \R^+$ and $\sigma(t)\in \{1,-1\}$ denote the position and velocity state of the RTP at time $t$. The corresponding velocity of the particle is $v(t)=v\sigma(t)$ for some constant speed $v>0$. Suppose that away from the wall the particle randomly switches its velocity state at the sequence of times ${\calT}_{n}$ generated from a homogeneous Poisson process $N(t)$ with a fixed rate $\alpha$. The Poisson process counts the number of tumbles occurring in the interval $[0,t]$. In other words,
\begin{eqnarray*}
N(t)=n,\quad  {\calT}_{n}\leq t < {\calT}_{n+1}.
\end{eqnarray*}
Note that $N(t)$ is defined to be right-continuous, $\lim_{\epsilon \rightarrow 0^+}N(t+\epsilon)=N(t)$. In particular, $N( {\calT}_n^-)=n-1$ whereas $N({\calT}_n)=n$. We assume that the particle is also subject to Gaussian white noise with a constant diffusion coefficient $D$. A non-trivial step is incorporating the effects of the wall on the SDE describing the evolution of $(X(t),\sigma(t))$. In previous work we solved this problem in the simpler case of an RTP without diffusion for both a non-sticky wall \cite{Bressloff25a} and a sticky wall \cite{Bressloff25b}. Here we extend the analysis to include the effects of diffusion. 
 
\subsection{Weighted local times and reflected motion at a hard wall}

Suppose for the moment that the boundary at $x=0$ is treated as a hard wall (no absorption). In the absence of diffusion ($D=0$), there are two distinct models of particle/wall interactions, see also Fig. \ref{fig1}. The first assumes that the wall is totally reflecting -- whenever the RTP collides with the wall, it instantaneously reverses its direction from $-v$ to $v$ and reenters the bulk domain. (The particle cannot hit the wall in the right-velocity state.) This means that wall collisions are discrete events that can be represented by a second counting process $L(t)$. The latter denotes the number of collisions over the time interval $[0,t]$. (Note that $L(t)$ is also right-continuous but is generally non-Poissonian.) Following Ref. \cite{Bressloff25a}, we set
\begin{equation}
\label{locmA}
L(t) =v \int_0^t\delta_{\sigma(\tau),-1}\delta(X(\tau))d\tau   =\int_0^t \sum_{n\geq 1}\delta(\tau-\calS_n)d\tau,
\end{equation}
where $\calS_n$ is the time of the $n$th collision with the wall. Each trajectory of the RTP will typically yield a different value of $L(t)$, which means that $L(t)$ is itself a stochastic process. 
The second model assumes that, rather than instantaneously reflecting, an incoming particle becomes stuck at the wall until it reverses direction from $-v$ to $v$. This is an example of a sticky boundary \cite{Angelani15,Angelani17,Bressloff23}. One way to model a sticky boundary is to introduce the probability $Q_0(t)$ that at time $t$ the particle is attached to the wall in the bound state $B_0$. Equations (\ref{KA0}) and (\ref{KB0}) are then supplemented by the boundary condition 
\begin{equation}
\label{mstick0}
vp_1(0,t) =\gamma Q_0(t),
\end{equation}
where $\gamma$ is the tumbling rate at the wall and $Q_0(t)$ evolves according to the equation
\begin{equation}
\label{stick0}
\frac{dQ_0}{dt}=vp_{-1}(0,t)-\gamma  Q_0(t) .
\end{equation}
Note that in the limit $\gamma \rightarrow \infty$, the particle returns to the bulk as soon as it hits the wall and we recover the totally reflecting boundary condition
$
p_1(0,t)=p_{-1}(0,t)$.
On the other hand, taking the limit $\gamma \rightarrow 0$ leads to the totally adsorbing boundary condition
$p_1(0,t)=0$.

\begin{figure}[t!]
\centering
\includegraphics[width=12cm]{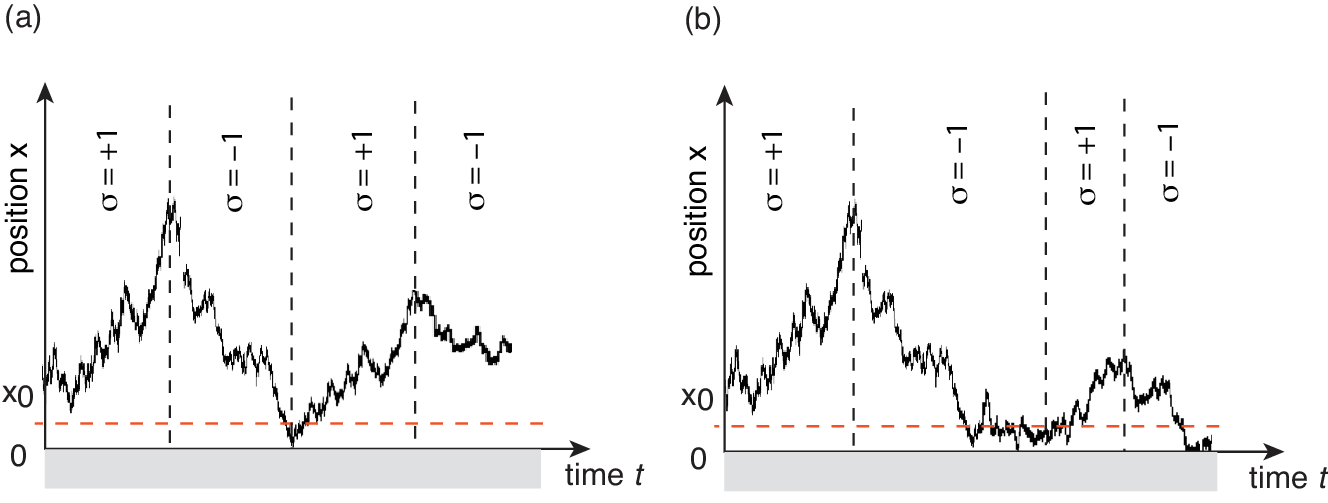} 
\caption{RTP with diffusion confined to the half-line with a hard wall at $x=0$. Horizontal dashed line indicates a boundary layer. (a) Schematic diagram illustrating a trajectory for model A. Each time the particle hits the wall in the $(-)$ state it immediately switches to the $(+)$ state. The particle then spends a short time within the boundary layer before quickly exiting due to the positive drift velocity. (b) Schematic diagram illustrating a trajectory for model B. Each time the particle hits the wall in the $(-)$ state it tends to spend significant time within the boundary layer until switching to the $(+)$ state.}
\label{fig2}
\end{figure}

The situation is very different when diffusion is included ($D>0$) since the RTP can hit the wall in either velocity state ($\sigma(t)=\pm 1$). Let us define the weighted local times
\begin{equation}
\ell_{\pm }(t)=D \int_0^t\delta_{\sigma(\tau),\pm 1}\delta(X(\tau))d\tau .
\end{equation}
The local time $\ell_{\pm 1}(t)$ keeps track of the amount of time the RTP spends in contact with the wall while in the state $\sigma=\pm 1$. Note that $\ell_+(t)$ is a continuous stochastic process, which is a weighted version of the boundary local time for a Brownian particle with constant drift $v>0$ \cite{Grebenkov22}. On the other hand, $\ell_-(t)$ can be a discrete or continuous stochastic process, depending on whether drift velocity reversals ($-v\rightarrow v$) occur when the particle collides with the wall in the $(-)$ state. If reversals do occur then $\ell_-(t)$ is discrete and is related to the counting process $L(t)$ according to $\ell_{-}(t)=DL(t)/v$. Conversely, if there are no reversals, then $\ell_-(t)$ is a weighted version of the continuous boundary local time for a Brownian particle with constant drift $v<0$. We will refer to these two models of reflected motion as model A and model B, respectively. In Fig. \ref{fig2} we illustrate the differences between the two models with regards reflected motion. It can be seen that the amount of time spent in a boundary layer close to the wall is greatly enhanced in model B, and these periods of close contact occur whenever the particle hits the wall in the $(-)$ state. Such periods are analogous to being attached to a sticky wall in the limit $D\rightarrow 0$.

 We represent the stochastic single-particle dynamics for $D>0$ by combining the SDE of reflected Brownian motion with the SDE of a non-diffusing RTP, which was introduced in Ref. \cite{Bressloff25a}. More specifically, we take
\numparts
\begin{eqnarray}
\label{dX}
dX(t)&=v\sigma(t)dt+\sqrt{2D}dW(t) +d\ell(t),\\
d\sigma(t)&=-2  \sigma(t^-)[dN(t)+\Gamma_0 dL(t)],
\label{dsig}
\end{eqnarray}
\endnumparts
where $W(t)$ is pure Brownian motion, $\ell(t)=\ell_+(t)+\ell_-(t)$, and
\numparts
\begin{eqnarray}
\label{ht}
dN(t)&=\chi(t)dt,\quad \chi(t)=\sum_n\delta (t-T_n),\\
 \label{PDMPb}
dL(t)&= \eta(t) dt,\quad \eta(t)=v\delta(X(t))\delta_{\sigma(t),-1},\\
 d\ell(t)&= D\delta(X(t)).
 \end{eqnarray}
 \endnumparts
 The constant $\Gamma_0$ in equation (\ref{dsig}) takes the value $\Gamma_0=1$ for model A and $\Gamma_0=0$ for model B. This ensures that, in the former case, whenever a particle in the $(-1)$ state collides with the wall it immediately reverses its internal velocity state, consistent with reflected run-and-tumble motion without diffusion.
The differential $d\ell(t)$ on the right-hand side of equation (\ref{dX}) ensures that whenever the particle hits the wall, it receives an impulsive kick in the positive $x$-direction, consistent with reflected Brownian motion. (Indeed, if $v=0$ for all $t$, then (\ref{dX}) reduces to the Skorokhod equation for reflected Brownian motion \cite{McKean75}.)

\subsection{Partially absorbing wall and local time thresholds}

Now suppose that the wall is partially absorbing. Following previous encounter-based models of diffusion \cite{Grebenkov06,Grebenkov20,Bressloff22} and RTPs without diffusion \cite{Bressloff22b,Bressloff23}, we incorporate partial absorption into the SDE of an RTP with diffusion, see equations (\ref{dX}) and (\ref{dsig}), by introducing a quenched random threshold $h$. In the case of model A, we assume that the particle is absorbed by the wall when the number of collisions $L(t)$ exceeds a discrete threshold $h$. This means that collisions with the wall when the particle is in the $(+)$ state are discounted with respect to the absorption process. The rule for partial absorption is illustrated in Fig. \ref{fig1} for a spatially discrete version of the model. The corresponding stopping time or first-passage time for absorption is then defined as
\begin{equation}
\label{calTA}
\fT=\int\{t>0:\, L(t) >\well\}.
\end{equation}
In section 3.1 we will show that if $h$ is generated from a geometrical distribution, then the boundary condition of the corresponding Kolmogorov equation is of Robin type. Conversely, in model B we assume that the particle is absorbed when the total local time $\ell(t)$ exceeds a continuous threshold $h$ so that
\begin{equation}
\label{calTB}
\fT=\int\{t>0:\, \ell(t) >\well\}.
\end{equation}
If $h$ is generated from an exponential distribution, then the resulting boundary conditions of the forward Kolmogorov equation are of Robin type, and are a generalisation of the totally reflecting or absorbing boundary conditions assumed in \cite{Malakar18}, see section 3.2.

\section{Derivation of the forward Kolmogorov equation} 
In this section we use the stochastic calculus of jump-diffusion processes and a generalised It\^o's lemma to derive the forward Kolmogorov equation for the probability densities $p_k(x,t)$, $k=\pm 1$.

\subsection{Model A ($\Gamma_0=1$)}

 Introduce the pair of $\well$-dependent empirical measures 
\begin{eqnarray}
\rho_{k,\well}(x,t)= \delta(x-X(t)){\bf 1}_{L(t)\leq \well} \delta_{k,\sigma(t)},\quad k=\pm 1,
\end{eqnarray} 
where ${\bf 1}_{L(t)\leq \well}=1$ if $L(t)\leq h$ and is zero otherwise. The latter factor incorporates the constraint that the particle hasn't yet been absorbed up to time $t$.
Consider a corresponding pair of arbitrary smooth test functions $f(x,k)$, $k=\pm 1$,  of compact support and set
\begin{equation}
\calF(t)=f(X(t),\sigma(t)){\bf 1}_{L(t)\leq \well}=\sum_{k\pm 1}\int_0^{\infty}  f(x,k)\rho_{k,\well}(x,t)dx.
\end{equation}
Differentiating both sides with respect to $t$ gives
\begin{eqnarray}
 &\sum_{k=\pm 1}\left [ \int_{\R^+}  f(x,k)\frac{\partial \rho_{k,\well}(x,t)}{\partial t}dx \right ] =\frac{d\calF(t)}{dt}  .\label{fluff0}
\end{eqnarray}
Applying It\^o's formula for  jump-diffusion processes to evaluate $d\calF(t)$, we have
\begin{eqnarray}
\label{Ito} 
 \fl   d\calF(t)  &={\bf 1}_{L(t)\leq \well}\bigg \{ \bigg [ D\delta(X(t))\partial_xf(X(t),\sigma(t))+\L_{\sigma}f(X(t),\sigma(t)) \bigg ]dt  \\
\fl & \quad +\sqrt{2D} \partial_xf(X(t),\sigma(t)) dW(t)  +\bigg [f(X(t),-\sigma(t))-f(X(t),\sigma(t))\bigg ]dN(t)\nonumber \\
 \fl &\quad +v\bigg [f(0,1) -f(0,-1)\bigg ]\delta(X(t)) \delta_{\sigma(t),-1}  \bigg \} - v\delta_{L(t),h} f(0,1)\delta(X(t))\delta_{\sigma(t),-1} .\nonumber
  \end{eqnarray}
  We have introdced the infinitesimal generator
  \begin{equation}
  \L_{\sigma}=v\sigma \frac{\partial}{\partial x}+D\frac{\partial^2}{\partial x^2}.
  \end{equation}
  The penultimate line of equation (\ref{Ito}) represents the velocity reversal due to a collision with the boundary in the $(-)$ state and the final line ensures that the particle is absorbed when $L(t)=h$.
  Substituting equation (\ref{Ito}) into the right-hand side of (\ref{fluff0}) and using the definition of the empirical measure $\rho_{k,\well}$, we have 
 \begin{eqnarray}
 \fl  & \sum_{k=\pm 1}   \int_0^{\infty} f(x,k)\frac{\partial \rho_{k,\well}(x,t)}{\partial t}  dx
\nonumber \\
  \fl &= \sum_{k=\pm 1}  \int_{0} ^{\infty}\rho_{k,\well}(x,t) \bigg [D\delta(x)\partial_x f(x, k)   +\L_kf(x, k)+\sqrt{2D}\partial_x f(x, k) \xi(t)   \bigg ]dx 
   \nonumber \\
  \fl  &\quad +\chi(t)   \sum_{k=\pm 1}  \int_0^{\infty}   \rho_{k,\well}(x,t)  [f(x,-k)-f(x,k)  ] dx  \label{V2} \\
 \fl  &\quad  + v    \bigg [f(0,1)-f(0,-1)\bigg ]\rho_{-1,\well}(0,t)-v \delta_{L(t),h} f(0,1)\delta(X(t))\delta_{\sigma(t),-1}.\nonumber
\end{eqnarray}
We have set $dW(t)=\xi(t)dt$ where $\xi(t)$ is a Gaussian white noise process.

Performing an integration by parts and resumming then gives
 \begin{eqnarray}
 \fl & \sum_{k=\pm 1}   \int_0^{\infty} f(x,k)\frac{\partial \rho_{k,\well}(x, t)}{\partial t}  dx=-  \sum_{k=\pm 1}  \int_{0} ^{\infty}f(x,k) \partial_x\calJ_{k,\well}(x,t)dx \nonumber \\
 \fl  &\quad -  \sum_{k=\pm 1}  f(0,k) \calJ_{k,\well}(0,t)+ \chi(t) \sum_{k=\pm 1}  \int_0^{\infty} f(x,k)  [\rho_{-k,\well}(x, t)-\rho_{k,\well}(x, t)\bigg ] dx  \nonumber\\
  \fl  &\quad  + v    \bigg [f(0,1) -f(0,-1)\bigg ]\rho_{-1,\well}(0,t)- \delta_{L(t),h} f(0,1)\delta(X(t))\delta_{\sigma(t),-1},
\end{eqnarray}
where we have introduced the stochastic probability fluxes
\begin{equation}
\calJ_{k,\well}(x,t)=vk \rho_{k,\well}(x, t) -D\partial_x \rho_{k,\well}(x, t)+\sqrt{2D}\xi(t)\rho_{k,\well}(x, t) .
\end{equation}
Note that boundary terms involving $\partial_xf(0,k)$ cancel.
Since $f(x,k)$ and $f(0,k)$ are arbitrary, it follows that $\rho_{k,\well}$ satisfies the SPDE (in the weak sense)
\numparts
\begin{eqnarray}
\fl  \frac{\partial \rho_{k,\well}(x,t)}{\partial t}&= - \frac{\partial}{\partial x} \calJ_{k,\well}(x,t) +\chi(t) [\rho_{-k,\well}(x, t^-)-\rho_{k,\well}(x, t^-)],\quad k=\pm 1,
  \label{0spuda}
  \end{eqnarray}
supplemented by the boundary conditions 
\begin{eqnarray}
 & D\frac{\partial}{\partial x} \rho_{1,\well}(0, t)-\sqrt{2D}\xi(t)\rho_{1,\well}(0, t) -v[\rho_{1,\well}(0,t)-\rho_{-1,\well}(0,t)] \nonumber \\
 &\hspace{2cm} =v \delta_{L(t),h}  \delta(X(t))\delta_{\sigma(t),-1},
\label{0spudb}  \\
 & D\frac{\partial}{\partial x} \rho_{-1,\well}(0, t)-\sqrt{2D}\xi(t)\rho_{-1,\well}(0, t) =0.
\label{0spudc}
\end{eqnarray}
\endnumparts
The final step in the derivation of the Kolmogorov equation is to average respect to the white noise process $\xi(t)$ and the Poisson noise process $N(t)$. Let
\begin{equation}
\label{defpk}
 p_{k,h}(x,t)=\bigg \langle \E[\rho_{k,h}(x,t)]\bigg \rangle,
\end{equation}
and
\begin{equation}
J_{k,h}(x,t)=\bigg \langle \E[\calJ_{k,h}(x,t)]\bigg \rangle =vkp_{k,h}(x,t)-D\frac{\partial p_{k,h}(x,t)}{\partial x},
\end{equation}
where $\E[\cdot ]$ and $\langle \cdot \rangle$ denotes expectations with respect to $\xi(t)$ and $N(t)$, respectively.
Taking expectations of equations (\ref{0spuda})--(\ref{0spudc}) then yields
\numparts
\begin{eqnarray}
\label{Aha}
&\frac{\partial p_{k,h}}{\partial t}=-v \frac{\partial p_{k,h}} {\partial x}+D \frac{\partial^2 p_{k,h}} {\partial x^2}+\alpha [p_{-k,h}-p_{k,h}],\quad x >0,
\end{eqnarray}
with the boundary conditions
\begin{eqnarray}
\fl & D\frac{\partial}{\partial x} p_{1,h}(0, t)  -v[p_{1,h}-p_{-1,h}(0,t)]  
 =v\bigg  \langle \E\bigg [\delta_{L(t),h}  \delta(X(t))\delta_{\sigma(t),-1} \bigg ]\bigg \rangle,
\label{Ahb}  \\
\fl  & D\frac{\partial}{\partial x}p_{-1,h}(0, t) =0.
\label{Ahc}
\end{eqnarray}
\endnumparts
We have used the following result for a homogeneous Poisson process:
\begin{equation}
\int_{\tau}^t\E[dN(s)] = \E[N(t)-N(\tau)]=\alpha (t-\tau),
\end{equation}
which implies that $\E[\chi(t)]=\alpha$.

Note that if $0
 < h <\infty$, then we do not have a closed system of equations for $p_{k,h}(x, t)$ due to the term on the right-hand side of the boundary condition (\ref{Ahb}). The two exceptions are $h=0$ (totally absorbing wall) and $h\rightarrow \infty$ (totally reflecting wall). In the first case, ${\bf 1}_{L(t)\leq 0}=\delta_{L(t),0}$ so that the right-hand side of (\ref{Ahb}) becomes $v\rho_{-1}(0,t)$, whereas in the second case the right-hand side vanishes. We thus obtain the closed Kolmogorov equation (after dropping the $h$-index)
 \begin{eqnarray}
\label{FKol}
&\frac{\partial p_{k}}{\partial t}=-v \frac{\partial p_{k}} {\partial x}+D \frac{\partial^2 p_{k}} {\partial x^2}+\alpha [p_{-k}-p_{k}],\quad x >0,
\end{eqnarray}
with either the totally reflecting boundary conditions 
\numparts
\begin{eqnarray}
\label{ref}
 \fl & -D \frac{\partial p_{1}(0,t)} {\partial x}+ vp_1(0,t)=vp_{-1}(0,t) ,\quad  D \frac{\partial p_{-1}(0,t)} {\partial x}=0 ,\end{eqnarray}
or the totally absorbing boundary conditions
\begin{eqnarray}
\label{abs}
 \fl & -D \frac{\partial p_{1}(0,t)} {\partial x}+ vp_1(0,t) =0,\quad  D \frac{\partial p_{-1}(0,t)} {\partial x}=0. \end{eqnarray}
 \endnumparts
The interpretation of the totally reflecting boundary conditions (\ref{ref}) is as follows: (i) the $(+)$ and $(-)$ boundary fluxes are equal due to instantaneous reflections and (ii) the $(-)$ boundary flux is purely ballistic. In the totally absorbing case, equations (\ref{abs}), the $(-)$ boundary flux is immediately absorbed rather than reflected so the $(+)$ boundary flux is zero. 
Following our previous work \cite{Bressloff25b,Bressloff25a}, there is another case where one can obtain a closed system of equations, but this requires averaging with respect to the quenched random threshold $h$. Denote the probability distribution  of $h$ by
 \begin{equation}
 \label{psi0}
 \psi(\ell)=\P[h=\ell],\quad \sum_{\ell=0}^{\infty}\psi(\ell)=1.
 \end{equation}
Also note the relations
 \begin{equation}
 \label{Psi0}
\fl \Psi(\ell) \equiv \P[h>\ell]=\sum_{\ell'=\ell+1}^{\infty} \psi(\ell')=1-\sum_{\ell'=0}^{\ell}\psi(\ell'),\quad \psi(\ell)=\Psi(\ell-1)-\Psi(\ell).
\end{equation}
We then define
\begin{eqnarray}
p_{k}(x,t)&=\sum_{h=0}^{\infty}\psi(h) p_{k,h}(x,t)   
\nonumber \\
&=\sum_{h=0}^{\infty}\psi(h) \bigg \langle \E \bigg [\delta(x-X(t)){\bf 1}_{L(t)\leq \well} \delta_{k,\sigma(t)} \bigg ]\bigg \rangle\nonumber \\
&=  \bigg\langle \E \bigg [\delta(x-X(t))\Psi(L(t)-1) \delta_{k,\sigma(t)} \bigg ]\bigg \rangle.
\end{eqnarray}
Multiplying both sides of equation (\ref{Aha}) by $\psi(h)$ and summing over $h$ then yields the Kolmogorov equation (\ref{FKol}) with the boundary conditions
\begin{eqnarray*}
\fl  & -D \frac{\partial p_{1}(0,t)} {\partial x}+ v[p_1(0,t)-p_{-1}(0,t)] = -v   \bigg\langle \E \bigg [\delta(x-X(t))\psi(L(t)) \delta_{-1,\sigma(t)} \bigg ]\bigg \rangle, \\
\fl &  D \frac{\partial p_{-1}(0,t)} {\partial x}=0 .
 \end{eqnarray*}
It is clear that for a general threshold distribution $\psi(h)$, the $(+)$ boundary condition is not closed. However, 
following Refs. \cite{Bressloff22b,Bressloff25a}, suppose that $\Psi(\ell)$ is the geometric distribution
\begin{equation}
\label{odd}
 \Psi(\ell)=z^{\ell+1}, \quad \psi(\ell)=(1-z)z^{\ell} , \quad \ell\geq 0,\quad z\in [0,1].
\end{equation}
We then have
\begin{eqnarray*}
\fl   \bigg\langle \E \bigg [\delta(x-X(t))\psi(L(t)) \delta_{-1,\sigma(t)} \bigg ]\bigg \rangle &=z\bigg\langle \E \bigg [\delta(x-X(t))z^{L(t)} \delta_{-1,\sigma(t)} \bigg ]\bigg \rangle\\
&=(1-z)p_{-1}(x,t).
\end{eqnarray*}
The $(+)$ boundary condition thus reduces to the closed ``Robin'' form so that equations  (\ref{FKol}) are supplemented by the closed boundary conditions
 \begin{equation}
 \label{BCmodelA}
 \fl -D \frac{\partial p_{1}(0,t)} {\partial x}+ v[p_1(0,t)-p_{-1}(0,t)] =-\kappa_0p _{-1}(0,t),\quad  D \frac{\partial p_{-1}(0,t)} {\partial x}=0,
\end{equation}
with the ``reactivity''
\begin{equation}
\label{kaz}
\quad \kappa_0=\kappa(z)\equiv v(1-z),\quad z\in [0,1].
\end{equation}

\subsection{Model B ($\Gamma_0=0$)} 

 In the case of model B we consider the pair of $\well$-dependent empirical measures 
\begin{eqnarray}
\rho_{k,\well}(x,t)= \delta(x-X(t)){\bf 1}_{\ell(t)\leq \well} \delta_{k,\sigma(t)},\quad k=\pm 1,
\end{eqnarray} 
where ${\bf 1}_{\ell(t)\leq \well}=\Theta(h-\ell(t)$ with $\ell(t)=\ell_+(t)+\ell_-(t)$ a continuous function of time $t$ and $h$ a continuously distributed quenched random variable. Since the basic steps in the derivation of the Kolmogorov equation are the same, we only indicated differences in the detailed equations. First, the It\^o formula (\ref{Ito}) becomes
\begin{eqnarray}
\label{ItoB} 
 \fl   d\calF(t)  &=\Theta(h-\ell(t))\bigg \{ \bigg [ D\delta(X(t))\partial_xf(X(t),\sigma(t))+\L_{\sigma}f(X(t),\sigma(t)) \bigg ]dt  \\
\fl & \quad +\sqrt{2D} \partial_xf(X(t),\sigma(t)) dW(t)  +\bigg [f(X(t),-\sigma(t))-f(X(t),\sigma(t))\bigg ]dN(t)\nonumber \\
 \fl &\quad -D\delta(h-\ell(t))f(0,\sigma(t))\delta(X(t)).\nonumber
  \end{eqnarray}
  The only difference is the last line on the right-hand side of equation (\ref{ItoB}).
 We now substitute equation (\ref{ItoB}) into the right-hand side of (\ref{fluff0}), use the definition of the empirical measure $\rho_{k,\well}$, and perform various integration by parts. Finally, exploiting the arbitrariness of the test function $f$ yields the SPDE (\ref{0spuda})
supplemented by the modified boundary conditions 
\numparts
\begin{eqnarray}
  & D\frac{\partial}{\partial x} \rho_{1,\well}(0, t)-\sqrt{2D}\xi(t)\rho_{1,\well}(0, t) -v\rho_{1,\well}(0,t) ,\nonumber \\\fl &\hspace{2cm}=D\delta(h-\ell(t)) \delta(X(t)),
\label{Bspudb}  \\
 & D\frac{\partial}{\partial x} \rho_{-1,\well}(0, t)-\sqrt{2D}\xi(t)\rho_{-1,\well}(0, t) -v\rho_{-1,\well}(0,t)\nonumber \\ \fl &\hspace{2cm} =D\delta(h-\ell(t)) \delta(X(t)).
\label{Bspudc}
\end{eqnarray}
\endnumparts
Similarly, averaging with respect to the white noise process $\xi(t)$ and the Poisson noise process $N(t)$ yields equation (\ref{Aha}) 
\numparts
with the modified boundary conditions
\begin{eqnarray}
\fl & D\frac{\partial}{\partial x} p_{1,h}(0, t)  -vp_{1,h} 
 =D\bigg  \langle \E\bigg [\delta(h-\ell(t)) \delta(X(t))  \bigg ]\bigg \rangle,
\label{Bhb}  \\
\fl  & D\frac{\partial}{\partial x}p_{-1,h}(0, t) +vp_{-1,h}(0, t) =D\bigg  \langle \E\bigg [\delta(h-\ell(t)) \delta(X(t))  \bigg ]\bigg \rangle.
\label{Bhc}
\end{eqnarray}
\endnumparts

Following the encounter-based model of diffusion-mediated absorption \cite{Grebenkov20,Bressloff22}, we now average with respect to the quenched random threshold $h$. Let $\Psi(r)=\P[h>r]$ with $\psi(r)=-\Psi'(r)$ and $\Psi(r)=\int_{r}^{\infty} \psi(r')dr'$. Define
\begin{eqnarray}
p_{k}(x,t)&=\int_{0}^{\infty}\psi(h) p_{k,h}(x,t)  dh 
\nonumber \\
&=\int_0^{\infty}\psi(h) \bigg \langle \E \bigg [\delta(x-X(t))\Theta(h-{\ell(t)} \delta_{k,\sigma(t)} \bigg ]\bigg \rangle dh\nonumber \\
&=  \bigg\langle \E \bigg [\delta(x-X(t))\Psi(\ell(t)) \delta_{k,\sigma(t)} \bigg ]\bigg \rangle.
\end{eqnarray}
Multiplying both sides of equations (\ref{Aha}), (\ref{Bhb}) and (\ref{Bhc}) by $\psi(h)$ and integrating with respect to $h$ then yields the Kolmogorov equation (\ref{FKol}) with
\numparts
\begin{eqnarray}
 & D\frac{\partial}{\partial x} p_{1}(0, t)  -vp_{1} 
 =D\bigg  \langle \E\bigg [\psi(\ell(t)) \delta(X(t))  \bigg ]\bigg \rangle,
\label{Bhb2}  \\
   & D\frac{\partial}{\partial x}p_{-1}(0, t) +vp_{-1}(0, t) =D\bigg  \langle \E\bigg [\psi(\ell(t)) \delta(X(t))  \bigg ]\bigg \rangle.
\label{Bhc2}
\end{eqnarray}
\endnumparts
For a general probability density $\psi(h)$ we do not have a closed system of equations. However, in the special case of the exponential distribution $\Psi\ell)=\e^{-\kappa_0\ell/D}$ we have $\psi(\ell)=\kappa_0\Psi(\ell)/D$. We thus obtain the closed boundary conditions
\begin{eqnarray}
  & -D\frac{\partial}{\partial x} p_{k}(0, t)  +vkp_{k}(0,t) 
= -\kappa_0 p(0,t),\quad k=\pm 1,
\label{BCmodelB}
\end{eqnarray}
where $p=p_1+p_{-1}$. Setting $\kappa_0=0$ recovers the boundary conditions of a totally reflecting hard wall assumed in Ref. \cite{Malakar18}. On the other hand, taking the limit $\kappa_0\rightarrow \infty$ yields the totally absorbing boundary conditions $p_{\pm 1}(0,t)=0$ (assuming $v>0$).

\subsection{Hard walls ($\kappa_0=0$) and the zero-diffusion limit}

In summary, for an RTP with diffusion and a partially absorbing wall of constant reactivity $\kappa_0$, the forward Kolmogorov equation is given by equation (\ref{FKol}) and either the boundary conditions (\ref{BCmodelA}) for model A or the boundary conditions (\ref{BCmodelB}) for model B. In order to explore how the two models differ in the zero-diffusion limit, we focus on the case of a hard wall ($\kappa_0=0$). First, taking the limit  $D\rightarrow 0$ in equations (\ref{BCmodelA}), we find that the $(-)$ boundary condition becomes an identity, whereas the $(+)$ boundary condition reduces to the familiar form \cite{Angelani15,Angelani17} $p_1(0,t)=p_{-1}(0,t)$. (Similarly, if $0<\kappa_0\leq v$ we obtain the Robin-like boundray condition derived in Ref. \cite{Bressloff22a}.) On the other hand, naively setting $\kappa_0=0$ and $D=0$ in equations (\ref{BCmodelB}) yields the pair of ``absorbing'' boundary conditions $p_{\pm 1}(0,t)=0$, which also means that the model is overdetermined. A classical method for dealing with this issue is to use singular perturbation theory, see Sect. 3.4 for details. The basic idea is to take $D=O(\epsilon)$ and to introduce a boundary layer of width $O(\epsilon)$ within which $\partial p_{-1}/\partial x=O(1/\epsilon)$. Hence, one cannot simply set $D\partial_xp_{-1}(0,t)=0$ in the limit $\epsilon \rightarrow 0$. One consequence of the boundary layer analysis is that
the probability $Q_{\epsilon}(t)=\int_0^{\epsilon} p_{-1}(x,t)dx$ remains finite as $\epsilon \rightarrow 0$, which is equivalent to having a sticky hard wall. Such a result can also be obtained from model B by taking the $D\rightarrow 0$ limit of the steady-state solution of an RTP with diffusion confined to a finite interval with hard walls at either end ($\kappa_0=0$) \cite{Malakar18}. Here we briefly extend this steady-state analysis to include model A.

Consider the steady-state Kolmogorov equation for an RTP in the interval $[-a,a]$ with a hard wall at both ends:
 \begin{eqnarray}
&0=-v \frac{dp_{k}}{d x}+D \frac{d^2 p_{k}} {dx^2}+\alpha [p_{-k}-p_{k}],\quad-a<x<a,
\end{eqnarray}
with
\begin{eqnarray}
 & -D \frac{d p_{1}}{d x}+ v[p_1-p_{-1}] =0,\quad  D \frac{d p_{-1}} {d x}=0,\quad x=\pm a \end{eqnarray}
 for model A and
 \begin{eqnarray}
 & -D \frac{d p_{1}}{d x}+ vp_1  =0,\quad  -D \frac{d p_{-1}} {d x}-vp_{-1} =0,\quad x=\pm a \end{eqnarray}
for model B.
 Introducing the change of variables $q=p_1-p_{-1}$ and $p=p_1+p_{-1}$ we have
\numparts
 \begin{eqnarray}
\label{ssFa}
&0=D \frac{d^2 p} {dx^2}-v \frac{dq}{d x}=0,\quad-a<x<a,\\
\label{ssFb}
&0=D \frac{d^2 q} {dx^2}-v \frac{dp}{d x}-2\alpha q,\quad-a<x<a,
\end{eqnarray}
with
\begin{eqnarray}
\label{ssFc}
 & -D \frac{d p }{d x}+ vq =0,\quad  -D \frac{d q} {d x}+vq=0,\quad x=\pm a\end{eqnarray}
 for model A and
 \begin{eqnarray}
\label{ssFd}
 & -D \frac{d p }{d x}+ vq =0,\quad  -D \frac{d q} {d x}+vp=0 ,\quad x=\pm a\end{eqnarray}
 \endnumparts
 for model B. Only the second set of boundary conditions differ between the two models
 
 Integrating equation (\ref{ssFa}) and using the first set of boundary conditions shows that $D\partial_xp=vq$ for all $x\in [-a,a]$. Eliminating $p$ from equation (\ref{ssFb}) then gives the second-order equation
\[D\frac{ d^2q}{dx^2}-(2\alpha+v^2/D)q=0.\]
This has the general solution
\begin{equation}
\label{qx}
q(x)=A\e^{\mu x}+B\e^{-\mu x},\quad -a<x<a,\quad \mu=\frac{\sqrt{2\alpha D+v^2}}{D}.
\end{equation}
Substituting the solution for $q$ into the relation $D\partial_xp=vq$ and integrating yields
\begin{equation}
\label{px}
p(x)=\frac{v}{D \mu}\bigg (A\e^{\mu x}-B\e^{-\mu x}\bigg )+C.
\end{equation}
The three constants $A,B,C$ are then determined from the second set of boundary conditions and the normalisation condition
\begin{equation}
\int_{-a}^a[p_1(x)+p_{-1}(x)]dx\equiv \int_{-a}^ap(x)dx=1.
\end{equation}
\medskip

\noindent \underline{Model A.}  Substituting the solution (\ref{qx}) into the second set of boundary conditions in (\ref{ssFc}) yields the pair of equations
\begin{equation*}
\fl (v-D\mu)A\e^{\mu a} +(v+D\mu)B\e^{-\mu a}=0,\quad (v-D\mu)A\e^{-\mu a} +(v+D\mu)B\e^{\mu a}=0.
\end{equation*}
These can only be satisfied if $A=B=0$. It immediately follows that $q(x)=0$ and $p(x)=C$. The normalisation condition thus yields the uniform solution (which holds for all $D\geq 0$)
\begin{equation}
p_1(x)=p_{-1}(x)=\frac{1}{4a} ,\quad -a\leq x \leq a.
\end{equation}
\medskip

\noindent \underline{Model B.}  As shown in Ref. \cite{Malakar18} for model B, the steady-state solution $p(x)$ is
\begin{equation}
\fl p(x)=\left [\frac{v^2}{\alpha}\frac{\tanh(\sqrt{2\alpha D+v^2}\, a/D)}{\sqrt{2\alpha D+v^2}}+2a\right ]^{-1}\left [\frac{v^2\cosh(\sqrt{2\alpha D+v^2}\, x/D)}{2\alpha D\cosh(\sqrt{2\alpha D+v^2}\, a/D)}+1\right ].
\end{equation}
Taking the limit $D\rightarrow 0$ then yields 
\begin{equation}
\label{delQ}
\lim_{D\rightarrow 0} p(x)=\frac{1}{2a+v/\alpha}\left [1+\frac{v}{\alpha}\frac{\delta (x-a)+\delta(x+a)}{2} \right ].
\end{equation}
The Dirac delta functions represent the probability measures associated with bound states in the case of sticky walls at $x=\pm a$ \cite{Angelani17}. That is, let $Q_{\pm }$ denote the steady state probability that the RTP is stuck to the wall at $x=\pm a$. Using steady-state versions of equations (\ref{mstick0}) and (\ref{stick0}) with $\gamma =\alpha$ (same tumbling rate at the walls as in the bulk), we have $\alpha Q_{\pm}=v p_0/2$ where $p_0$ is the steady-state probability in the bulk for $D=0$. It follows from unit normalisation of the total probability that
\[\frac{vp_0}{\alpha} +2a p_0=1.\]
Hence,
\begin{equation}
 p_0=\frac{1}{2a +v/\alpha},\quad Q_{\pm}=\frac{v}{2\alpha}p_0.
 \end{equation}
 The right-hand of equation (\ref{delQ}) is an equivalent way of representing the total probability measure.

 \subsection{Boundary-layer analysis} As we previously mentioned, yet another way to understand the singular limit $D\rightarrow 0$ of model B is to use singular perturbation theory. Set $D=\epsilon D_0$ with $0<\epsilon \ll 1$ and $D_0$ independent of $\epsilon$.
First consider the outer solution
\[p(x)\sim p_0(x)+\epsilon p_1(x)+\ldots,\quad q(x)\sim q_0(x)+\epsilon q_1(x)+\ldots,
\]
which is valid in the bulk of the domain $-a<x<a$. Substituting into equations (\ref{ssFa}) and (\ref{ssFb}) and collecting  $O(1)$ terms shows that
 \begin{eqnarray}
\label{out}
0=-v \frac{dq_0}{d x}=0,\quad 0=-v \frac{dp_0}{d x}-2\alpha q_0,\quad-a<x<a.
\end{eqnarray}
This has the general solution
\begin{equation}
q_0(x)=C_1,\quad p_0(x)=\frac{2\alpha C_1}{v}x+C_2,
\end{equation}
where $C_1,C_2$ are integration constants. It is clear that the effective second-order equation is overdetermined by the pair of boundary conditions at $x=\pm a$. This can be dealt with by introducing a boundary layer at each end. Within the boundary layers, $d^2p/dx^2$ and $d^2q/dx^2$ vary rapidly so that a regular perturbation expansion breaks down.
 
 For the sake of illustration, consider the solution in a boundary layer around $x=-a$. Introduce the stretched coordinate $y=(x+a)/\epsilon$ so that equations (\ref{ssFa}), (\ref{ssFb}) and (\ref{ssFd}) become
 \numparts
 \begin{eqnarray}
\label{BLa}
&0=D_0 \frac{d^2 \widehat{p}} {dy^2}-v \frac{d\widehat{q}}{d y}=0,\quad 0<y<\infty,\\
\label{BLb}
&0=D_0 \frac{d^2 \widehat{q}} {dy^2}-v \frac{d\widehat{p}}{d y}-2\epsilon \alpha \widehat{q},\quad 0<y<\infty,
\end{eqnarray}
with
 \begin{eqnarray}
\label{BLc}
 & -D_0 \frac{d   \widehat{p}}{d y}+ v \widehat{q} =0,\quad  -D_0 \frac{d  \widehat{q}} {d y}+v \widehat{p}=0 ,\quad y=0.\end{eqnarray}
 \endnumparts
 The inner solution of equations (\ref{BLa})--(\ref{BLc}) then has to be matched with the outer solution in the bulk, that is, 
 \[\lim_{y\rightarrow \infty} \widehat{p}(y) =  \lim_{x\rightarrow 0} p(x),\quad \lim_{y\rightarrow \infty} \widehat{q}(y) = \lim_{x\rightarrow 0} q(x),\]
 where $p(x),q(x)$ satisfy the outer equations. Following along similar lines to the derivation of equations (\ref{qx}) and (\ref{px}), we find that imposing the first boundary condition in (\ref{BLc}) yields the solution
 \begin{equation}
 \fl \widehat{q}(y)\sim B'\e^{-\mu_0(\epsilon) y},\quad  \widehat{p}(y)\sim -\frac{vB'}{D \mu_0(\epsilon)}\e^{-\mu_0(\epsilon) y} +C',\quad 0<y<\infty,
 \end{equation}
 with
 \[\quad \mu_0(\epsilon)=\frac{\sqrt{2\alpha \epsilon D_0+v^2}}{D_0}.\]
 The second boundary condition then determines $B'$ in terms of $C'$ so that
 \begin{equation}
 \widehat{p}(y)\sim C' \bigg [1+\frac{v^2}{2\epsilon D_0 \alpha }\e^{-\mu_0(\epsilon) y}  \bigg ].
  \end{equation}
  The leading order version of the inner solution can be rewritten as
  \begin{equation}
 \widehat{p} \sim p_0\bigg [1+\frac{v^2}{2D_0\epsilon  \alpha }\e^{-v (x+a)/\epsilon D_0}  \bigg ].
  \end{equation}
  Finally, we note that
  \begin{equation}
\frac{v^2}{2\epsilon D_0  \alpha }\e^{-v (x+a)/\epsilon D_0}=\frac{v}{2\alpha}\delta_{\epsilon}(x+a),
\end{equation}
with 
\[\lim_{\epsilon \rightarrow 0} \delta_{\epsilon}(x) =\delta(x)\]
in the distribution sense.  
  
\section{First passage time densities on the half-line}

In Ref. \cite{Malakar18}, the first passage time (FTP) density of an RTP on the half line was analysed in the case of a totally absorbing wall. The authors proceeded by solving a backward Kolmogorov equation for the survival probability $S_k(y,t)$ that a particle starting in the state $(x(0),\sigma(0))=(y,k)$ has not been absorbed over the interval $[0,t]$. The corresponding boundary conditions were taken to be $S_{\pm 1}(0,t)=0$ for $D>0$ and $S_-(0,t)$ for $D=0$. Note that the FPT density $f_k(y,t)$  is related to $S_k(y,t)$ according to 
 \begin{equation}
 \label{fk}
 f_k(y,t)=-\frac{\partial S_k(y,t)}{\partial t}.
 \end{equation}
 Here we derive the backward Kolmogorov equation for the survival probability in the case of a partially absorbing wall and show that the adjoint boundary conditions for models A and B differ from those assumed in Ref. \cite{Malakar18}. The two models map, respectively, to the backward equations for non-sticky and sticky partially absorbing walls in the zero-diffusion limit.
 
 \subsection{Derivation of backward Kolmogorov equation}

First note that the survival probability can be written as
\begin{equation}
S_k(y,t)=\sum_{j=\pm 1}\int_0^{\infty}P_{jk}(x,t|y,0)dx,
\end{equation}
where $P_{jk}$ is the transition density or propagator
\begin{equation}
\label{tran}
\fl P_{jk}(x,t|x_0,t_0)dx=\P[x<X(t)<x+dx,\sigma(t)=j|X(t_0)=x_0,\sigma(t_0)=k].
\end{equation}
The transition densities satisfy the Chapman-Kolmogorov (CK) equation
\begin{eqnarray}
\label{ChapK}
P_{\sigma_1\sigma_0}(x,t|x_0,0)=\sum_{k=\pm 1} \int_0^{\infty} dy\, P_{\sigma k}(x,t|y,\tau)P_{k\sigma_0}(y,\tau|x_0,0).
\end{eqnarray}
This essentially states that the probability of being in the state $(x,\sigma)$ at time $t$ given the initial state $(x_0,\sigma_0)$ is the sum of the probabilities of all possible paths from $(x_0,\sigma_0)$ to $(x,\sigma)$ passing through an arbitrary intermediate point $(y,k)$ at time $\tau$, $0<\tau <t$. We will assume time translation invariance so that $P_{kk'}(x,t|y,\tau)=P_{kk'}(x,t-\tau|y,0)$.
Differentiating both sides of equation (\ref{ChapK}) with respect to the intermediate time $\tau$ gives
\begin{eqnarray*}
\fl 0&=\sum_{k=\pm 1} \int_{0}^{\infty}\partial_{\tau}P_{\sigma k}(x,t|y,\tau)P_{k\sigma_0}(y,\tau|x_0,0)dy +\int_{0}^{\infty}P_{\sigma k}(x,t|y,\tau)\partial_{\tau}P_{\sigma k}(y,\tau|x_0,0)dy .
\end{eqnarray*}
Using the fact that $p_k(y,\tau)=P_{k\sigma_0}(y,\tau|x_0,0)$ satisfies the forward Kolmogorov equation (\ref{FKol}), $\partial_{\tau}P_{k\sigma_0}(y,\tau|x_0,0)$ can be replaced by terms involving derivatives of $p_k(y,\tau)$
with respect to $y$:
\begin{eqnarray*}
\fl &0=\sum_{k=\pm 1}\int_{0}^{\infty}dy \bigg [\partial_{\tau}q_k(y,t-\tau)p(y,\tau)\nonumber \\
\fl &\qquad +q_k(y,t-\tau) \bigg (-kv\partial_{y}p_k(y,\tau) +D\partial^2_{yy}p_k(y, \tau) +\alpha [p_{-k}(y,\tau)-p_k(y,\tau)]\bigg ) , \end{eqnarray*}
where we have set $q_k(y,t-\tau)= P_{\sigma k}(x,t |y,\tau)$.
Integrating by parts with respect to $y$ then gives
\begin{eqnarray}
\fl &0=\sum_{k=\pm 1}\int_{0}^{\infty}\bigg[\partial_{\tau}q_k(y,t-\tau)+kv \partial_{y}q(y,t-\tau) +D\partial^2_{yy}q(y,t-\tau)\nonumber \\
\label{bac}
\fl &\hspace{2cm}+\alpha [q_{-k}(y,t-\tau)-q_k(y,t-\tau)]\bigg ]p(y,\tau) dy \\
\fl & -\left .\left [-kvq_k(y,t-\tau)p_k(y,\tau) +Dq_k(y,t-\tau)\partial_{y}p_k(y,\tau)-D\partial_{y}q_k(y,t-\tau)p_k(y,\tau)\right ]\right |_{y=0}.\nonumber
\end{eqnarray}
The integrand yields the backward Kolmogorov equation (after the change of time coordinates $\tau \rightarrow t-\tau $)
 \begin{eqnarray}
&\frac{\partial q_{k}}{\partial t}=vk \frac{\partial q_{k}} {\partial y}+D\frac{\partial^2 q_k}{\partial y^2}+\alpha [q_{-k}-q_{k}],\quad y >0.
\end{eqnarray}
The adjoint boundary conditions are then chosen so that the final line of equation (\ref{bac}) vanishes. In the case of model A, we substitute the forward boundary conditions (\ref{BCmodelA}) into equation (\ref{bac}), which implies that
\begin{eqnarray}
(\kappa_0-v)p_{-1}q_1+vp_{-1}q_{-1} -Dp_1 \partial_y q_1 -Dp_{-1}\partial_yq_{-1}=0
\end{eqnarray}
at $y=0$. This is satisfied if we impose the following adjoint boundary conditions for model A:
 \begin{equation}
 \fl D \frac{\partial q_{1}(0,t)} {\partial y}=0,\quad -D \frac{\partial q_{-1}(0,t)} {\partial y}+ v[q_1(0,t)-q_{-1}(0,t)] =\kappa_0q _{1}(0,t).
\end{equation}
Similarly, substituting the forward boundary conditions (\ref{BCmodelB}) into equation (\ref{bac}) leads to the following adjoint boundary conditions for model B:
 \begin{equation}
 \fl D \frac{\partial q_{1}(0,t)} {\partial y}=\kappa_0q (0,t),\quad D \frac{\partial q_{-1}(0,t)} {\partial y} =\kappa_0q (0,t),
\end{equation}
with $q=q_1+q_{-1}$. Finally, setting $q_k(y,t)= P_{\sigma k}(x,t |y,0)$, integrating with respect to $x$ and summing over $\sigma$ yields the following backward Kolmogorov equation for the survival probability $S_k$:
\begin{eqnarray}
\label{BKol}
&\frac{\partial S_{k}}{\partial t}=v \frac{\partial S_{k}} {\partial x}+D\frac{\partial^2 S_k}{\partial x^2}+\alpha [S_{-k}-S_{k}],\quad y >0,
\end{eqnarray}
together with the boundary conditions
 \begin{equation}
 \label{BBCmodelA}
 \fl D \frac{\partial S_{1}(0,t)} {\partial y}=0,\quad -D \frac{\partial S_{-1}(0,t)} {\partial y}+ v[S_1(0,t)-S_{-1}(0,t)] =\kappa_0S _{1}(0,t)
\end{equation}
for model A, and
 \begin{equation}
 \label{BBCmodelB}
 \fl D \frac{\partial S_{1}(0,t)} {\partial y}=\kappa_0S (0,t),\quad D \frac{\partial S_{-1}(0,t)} {\partial y} =\kappa_0S (0,t)
\end{equation}
for model B with $S=S_1+S_{-1}$. As with the forward Kolmogorov equation, model A is non-singular in the zero-diffusion limit since the first boundary condition in (\ref{BBCmodelA}) reduces to an identity. On the other hand, setting $D=0$ in equations (\ref{BBCmodelB}) yields an overdetermined system with the pair of boundary conditions $S_{\pm 1}(0,t)=0$. Again this would need to be handled by introducing a boundary layer or considering a partially absorbing sticky wall. We avoid these additional complexities by focusing here on the FPT densities for model A.

 \subsection{FPT densities for model A} 
 
 We begin by solving the backward Kolmogorov equations (\ref{BKol}) and (\ref{BBCmodelA}) for the survival probability in Laplace space. Using the initial conditions $S_{\pm 1}(y,0)=1$ and setting $\S_k(y,s)=\int_0^{\infty}\e^{-st}S_k(y,t)dt$, we have
 \numparts
 \begin{eqnarray}
\label{LTBKol0}
&-1+s\S_{k} =v \frac{\partial \S_{k}} {\partial x}+D\frac{\partial^2 \S_k}{\partial x^2}+\alpha [\S_{-k}-\S_{k}],\quad y >0,
\end{eqnarray}
and
 \begin{equation}
 \label{LTBBCmodelA}
 \fl D \frac{\partial \S_{1}(0,s)} {\partial y}=0,\quad -D \frac{\partial \S_{-1}(0,s)} {\partial y}+ v[\S_1(0,s)-\S_{-1}(0,s)] =\kappa_0\S _{1}(0,s).
\end{equation}
\endnumparts
Following Ref. \cite{Malakar18}, we can simplify equation (\ref{LTBKol0}) by performing the shift $\S_k(y,t)=s^{-1}+U_k(y,s)$ to give
\numparts
\begin{eqnarray}
\label{OO4a}
& \bigg [D\partial_y^2+v\partial_y-(\alpha+s)\bigg ]U_1(y,s)=-\alpha  U_{-1}(y,s),\\
&\bigg [D\partial_y^2-v\partial_y-(\alpha+s)\bigg ]U_{-1}(y,s)=-\alpha U_1(y,s),\quad y >0,
\label{OO4b}
\end{eqnarray}
and
 \begin{equation}
 \label{LTBBC}
 \fl D \frac{\partial U_{1}(0,s)} {\partial y}=0,\quad -D \frac{\partial U_{-1}(0,s)} {\partial y}+ v[U_1(0,s)-U_{-1}(0,s)] =\frac{\kappa_0}{s}+\kappa_0U_{1}(0,s).
\end{equation}
\endnumparts

Differentiating twice then yields separate closed equations for $U_{\pm}$:
\begin{equation}
\fl \bigg [D\partial_y^2+v\partial_y-(\alpha+s)\bigg ]\bigg [D\partial_y^2-v\partial_y-(\alpha+s)\bigg ]U_k(y,s)=\alpha^2 U_k(y,s),\quad y >0.
\label{O4}
\end{equation}
The bounded general solution is
\begin{equation}
U_k(y,s)= A_k(s)\e^{-\lambda_-(s) y/v}+B_k(s)\e^{-\lambda_+(s) y/v},
\end{equation}
where $\lambda_{\pm}(s)$ are the positive roots of the quartic characteristic equation
\begin{equation}
\Gamma_+(\lambda)\Gamma_-(\lambda)=\alpha^2 ,\quad \Gamma_{\pm}(\lambda)=\frac{D\lambda^2}{v^2}\pm  \lambda -(\alpha+s).
\end{equation}
That is, 
\begin{equation}
\fl \lambda_{\pm}(s)=v\left [\frac{v^2+2D(\alpha+s)\pm \sqrt{v^4+4v^2D(\alpha +s)+4\alpha^2 D^2}}{2D^2}\right ]^{1/2}.
\end{equation}
Requiring that the second-order equations (\ref{OO4a}) and (\ref{OO4b}) are also satisfied yields the relations
\begin{eqnarray}
\label{conA}
A_1&=-\frac{\Gamma_+(\lambda_-(s)) }{\alpha}A_{-1},\quad  B_1=-\frac{\Gamma_+(\lambda_+(s)) }{\alpha}B_{-1}.
\end{eqnarray}
 The remaining pair of equations required to determine the unknown coefficients are obtained by imposing the boundary conditions (\ref{LTBBC}). This gives
 \numparts
\begin{eqnarray}
\fl &\lambda_-(s)A_1+\lambda_+(s)B_1=0,\\
\fl &\frac{D}{v}\bigg [\lambda_-(s)A_{-1}+\lambda_-(s)B_{-1}\bigg ]+v\bigg [A_1+B_1-A_{-1}-B_{-1}\bigg ]=\frac{\kappa_0}{s}+\kappa_0\bigg [A_1+B_1\bigg ].\nonumber \\
\fl
\end{eqnarray}
\endnumparts
Using equations (\ref{conA}) to eliminate $A_1$ and $B_1$ gives
\numparts
\begin{eqnarray}
B_{-1}=-\frac{\lambda_-\Gamma_+(\lambda_-)}{\lambda_+\Gamma_+(\lambda_+)}A_{-1},
\end{eqnarray}
and
\begin{eqnarray}
  &\left [\frac{D\lambda_-}{v}-v\right ]A_{-1}+\left [\frac{D\lambda_+}{v}-v\right ]B_{-1}\nonumber \\
  &
=-\frac{\kappa_0-v}{\alpha}\bigg [\Gamma_+(\lambda_-)A_{-1}+\Gamma_+(\lambda_+)B_{-1}\bigg ]+\frac{\kappa_0}{s}.
\end{eqnarray}
\endnumparts
Combining the above two equations yields the result
\begin{eqnarray}
\fl A_{-1}&=\frac{\kappa_0}{s} \bigg \{\frac{\kappa_0-v}{\alpha}\Gamma_+(\lambda_-(s))\bigg (1-\frac{\lambda_-}{\lambda_+}\bigg )+\frac{D\lambda_-}{v}-v-\left [\frac{D\lambda_+}{v}-v\right ]  \frac{\lambda_-\Gamma_+(\lambda_-)}{\lambda_+\Gamma_+(\lambda_+)} \bigg \}^{-1}.\nonumber \\
\fl
\end{eqnarray}
Finally, Laplace transforming equation (\ref{fk}) implies that 
\begin{eqnarray}
\widetilde{f}_k(y,s)&=1-s\S_k(y,s)=-sU_k(y,s)\nonumber \\
&=-s\bigg [A_k(s)\e^{-\lambda_-(s) y/v}+B_k(s)\e^{-\lambda_+(s) y/v}\bigg ].
\label{solA}
\end{eqnarray}

\paragraph{Large time asymptotics.} Following Ref. \cite{Malakar18}, suppose that $s\rightarrow 0$ and $y\rightarrow \infty$ with $\sqrt{s}y$ fixed. This is equivalent to keeping $y/\sqrt{t}$ fixed. To leading order in $s$ we have
\[\lambda_-(s)\rightarrow v\sqrt{\frac{2\alpha s}{v^2+2\alpha D}},\quad \lambda_+(s)\rightarrow v\frac{\sqrt{v^2+2D\alpha}}{D}.
\]
Moreover, $A_{\pm 1}(s)\rightarrow -1/s$ and $sB_{\pm 1}\rightarrow O(\sqrt{s})$.
Taking the corresponding limit of $\widetilde{f}_k$ then gives
\begin{eqnarray}
\widetilde{f}_{\pm 1 }(y,s)&\rightarrow \e^{-y\sqrt{s/\calD }}
\end{eqnarray}
for all $0<\kappa_0 \leq v$, and thus
\begin{equation}
f_k(x,t)\rightarrow \frac{y}{\sqrt{4\pi  \calD t^3}}\e^{-y^2/4\calD t},
\end{equation}
where $\calD=D+v^2/2\alpha$. (The term $v^2/2\alpha$ is the effective diffusivity of the telegrapher equation in the so-called diffusion limit.) We deduce that the asymptotic behaviour of the FPT density $f_k$ is independent of the absorption rate $\kappa_0$ (for $0<\kappa_0\leq v$) and the initial direction of motion.

\subsection{Limit $D\rightarrow 0$.} Since model A is non-singular in the zero-diffusion limit, we can safely take the limit $D\rightarrow 0$ in equation (\ref{solA}). First, note that $\Gamma_{\pm}(s)= \pm \lambda -(\alpha+s)$ when $D=0$, and
\[\lim_{D\rightarrow 0} \lambda_-(s)=\lambda(s) \equiv \sqrt{ 2\alpha s+s^2 },\quad \lim_{D\rightarrow 0} D\lambda_+(s) =v^2.\]
Moreover $B_{\pm 1}\rightarrow 0$.
We thus obtain the following bounded solution for $D=0$:
\begin{equation}
U_k(y,s)= A_k(s)\e^{-\lambda(s) y/v},
\end{equation}
with
\numparts
\begin{equation}
A_{1}(s)=\frac{\alpha A_{-1}(s)}{\alpha +s+\lambda(s)},
\end{equation}
and
\begin{equation}
\fl A_{-1}(s)=\frac{\kappa_0}{s}\left [\frac{\alpha(v-\kappa_0)}{\alpha+s+v\lambda(s)}-v\right ]^{-1}=-\frac{\kappa_0}{s}\frac{\alpha+s+\lambda(s)}{\alpha \kappa_0+v(s+\lambda(s))}.
\end{equation}
\endnumparts
Note that in the totally absorbing case ($\kappa_0=v$), $A_{-1}(s)\rightarrow -1/s$.
Finally using the identity $\widetilde{f}_k(y,s)=-sU_k(y,s)$ we have
\numparts
\begin{eqnarray}
\label{f1p}
 \widetilde{f}_1(y,s)&=\frac{\kappa_0 \alpha }{\alpha \kappa_0+v(\lambda(s)+s)}\e^{-\lambda(s) y/v} ,\\
  \widetilde{f}_{-1}(y,s)&=\frac{\kappa_0 [\alpha+s+\lambda(s)]}{\alpha \kappa_0+v(s+\lambda(s))}\e^{-\lambda(s) y/v}.
\label{f1m}
\end{eqnarray}
\endnumparts

It is possible to obtain exact expressions for $f_k(y,t)$ by extending the analysis of Ref. \cite{Malakar18}. First, we define
\begin{equation}
\widetilde{g}_{\zeta}(y,s)=\widetilde{g}_{\zeta}(s)\e^{-\lambda(s)y/v},
\end{equation}
with
\begin{equation}
\label{gz}
\widetilde{g}_{\zeta}(s)=\frac{1}{\lambda(s)}\frac{1}{\kappa_0\alpha/ v+s+\lambda(s)}= \frac{1}{\lambda(s)}\frac{1}{\alpha+s+\lambda(s)-\alpha  \zeta},
\end{equation}
where $\zeta =1-\kappa_0/v$ and $0\leq \zeta <1$,
and rewrite equations (\ref{f1p}) and (\ref{f1m}) in the form
\numparts
\begin{eqnarray}
\label{f1p2}
\widetilde{f}_1(y,s)&=\alpha(1-\zeta) \lambda(s) \widetilde{g}_{\zeta}(y,s)=- \alpha(1-\zeta) v \partial_y \bigg [\widetilde{g}_{\zeta}(y,s)\bigg ],\\
\widetilde{f}_{-1}(y,s)&=\frac{\kappa_0}{v}\bigg [\e^{-\lambda(s) y/v}+\zeta \alpha \lambda(s)\widetilde{g}_{\zeta}(y,s) \bigg ]\nonumber \\
&=\frac{\kappa_0}{v}\bigg [\e^{-\lambda(s) y/v}+ \frac{  \zeta}{1-\zeta} \widetilde{f}_1(y,s)\bigg ]\nonumber \\
&=- \kappa_0 \partial_y  \bigg [\frac{\e^{-\lambda(s) y/v}}{\lambda(s)}\bigg ]+  \zeta  \widetilde{f}_1(y,s).
\label{f1m2}
\end{eqnarray}
\endnumparts
We can then invert the Laplace transforms by performing a Neumann series expansion of $\widetilde{g}_{\zeta}$ in powers of $\zeta$, and using the formulas
\numparts
\begin{eqnarray}
\fl  {\mathcal L}^{-1}\bigg [\widetilde{g}_0(y,s)\bigg ] &\equiv {\mathcal L}^{-1}\left [ \frac{\e^{-\lambda(s)y/v}}{\lambda(s)[s+\alpha +\lambda(s)]}\right ]\nonumber \\
\fl  & =\alpha^{-1}\e^{-\alpha t}\frac{\sqrt{t-y/v}}{\sqrt{t+y/v}}I_1(\alpha \sqrt{t^2-y^2/v^2})\Theta(t-y/v),
\label{LT1}
\end{eqnarray}
and
\begin{equation}
{\mathcal L}^{-1}\left [ \frac{\e^{-\lambda(s)y/v}}{\lambda(s)}\right ]=\e^{-\alpha t}I_0(\alpha \sqrt{t^2-y^2/v^2}))\Theta(t-y/v).
\label{LT2}
\end{equation}
\endnumparts
Here $\Theta(z)$ is a Heaviside function and $I_n(z)$ denotes a modified Bessel function of the first kind and integer order $n$. First, we rewrite equation (\ref{gz}) as
\begin{eqnarray}
\widetilde{g}_{\zeta}(s)&=\frac{1}{\lambda(s)}\frac{1}{\alpha+s+\lambda(s) }\frac{1}{1-\frac{\displaystyle \alpha  \zeta}{\displaystyle \alpha+s+\lambda(s)}}\nonumber \\
&=\widetilde{g}_{0}(s)\left [1+\frac{\displaystyle \alpha  \zeta}{\displaystyle \alpha+s+\lambda(s)}+O(\zeta^2)\right ].
\end{eqnarray}
Multiplying both sides by $\e^{-\lambda(s)y/v}$ and inverting using the convolution theorem gives
\begin{eqnarray}
\label{Neum}
\fl g_{\zeta}(y,t)=g_0(y,t)+ \zeta \alpha \int_0^{t}g_0(y,t-\tau)    H_1(\tau) d\tau+O(\zeta^2),
\end{eqnarray} 
where
\begin{eqnarray}
  H_1(\tau)\equiv {\mathcal L}_{\tau}^{-1}\bigg [\frac{1}{\alpha+s+\lambda(s) }\bigg ] &=(\alpha \tau)^{-1}\e^{-\alpha t} I_1(\alpha \tau).
\end{eqnarray}
Second, inverting equations (\ref{f1p2}) and (\ref{f1m2}) gives 
\numparts
\begin{eqnarray}
\label{f1}
\fl f_1(y,t)&=-\alpha(1-\zeta) v \partial_yg_{\zeta}(y,t),\\
\fl f_{-1}(y,t)&=(1-\zeta)\e^{-\alpha t}\bigg [  \alpha \frac{y/v}{\sqrt{t^2-y^2/v^2}}I_1(\alpha\sqrt{ t^2-y^2/v^2})\Theta(t-y/v)+ \delta(t-y/v)\bigg ]\nonumber \\
\fl &\quad +  \zeta {f}_1(y,t).
\label{fm1}
\end{eqnarray}
\endnumparts
Finally, differentiating equations (\ref{LT1}) and (\ref{Neum}) with respect to $y$ yields
\begin{eqnarray}
\label{Neumy}
\fl \partial_yg_{\zeta}(y,t)=\partial_y g_0(y,t)+ \zeta \alpha \int_0^{t}\partial_y g_0(y,t-\tau)    H_1(\tau) d\tau+O(\zeta^2),
\end{eqnarray} 
with
\begin{eqnarray}
\fl -\partial_y g_{0}(y,t)
 &=\alpha^{-1}\e^{-\alpha t}\bigg [ \frac{\alpha}{v} \frac{y/v}{t+y/v}I_0(\alpha\sqrt{ t^2-y^2/v^2})\nonumber \\
\fl &\hspace{2cm} + \frac{1/v}{t+y/v}\frac{\sqrt{t-y/v}}{\sqrt{t+y/v}}I_1(\alpha \sqrt{t^2-y^2/v^2})\bigg ]\Theta(\tau-y/v).\label{gb1}
\end{eqnarray} 
We have used the modified Bessel function identities
\[I_0'(z)=I_1(z),\quad I_1'(z)=I_0(z)-z^{-1}I_1(z).\]

\begin{figure}[t!]
\centering
\includegraphics[width=12cm]{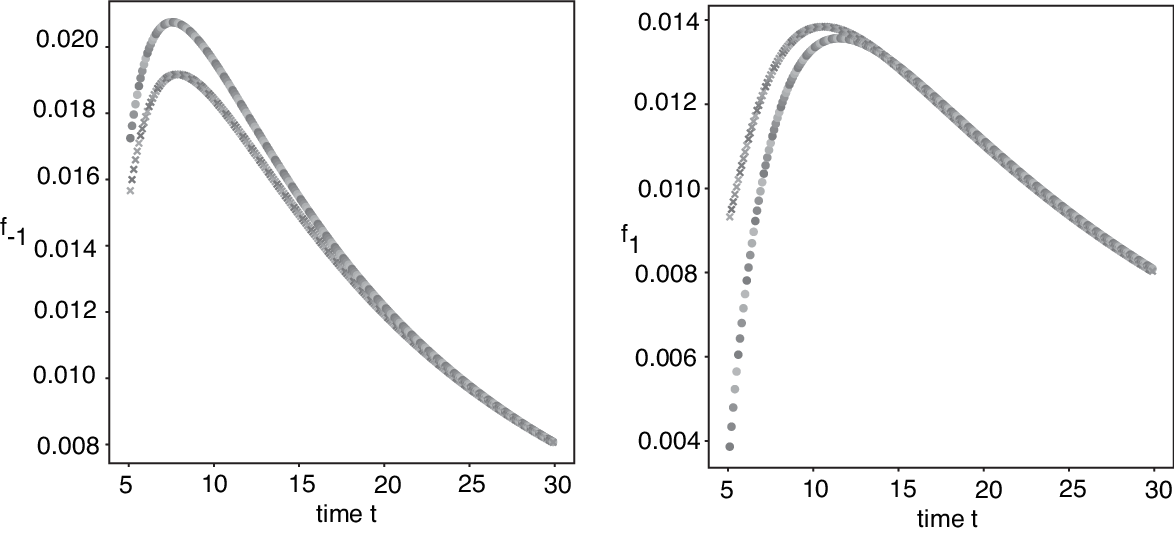} 
\caption{FPT densities for a partially absorbing non-sticky wall and $D=0$. Plots of (a) $f_{-1}(x_0,t)$ and (b) $f_1(x_0,t)$ as a function of time $t$ for fixed initial position $x_0=5$. Filled circles correspond to the case of a totally absorbing wall ($\zeta=0$), see equations (\ref{f10}) and (\ref{fm10}). Crosses represent the solutions (\ref{f1}) and (\ref{fm1}) based on the first two terms in the Neumann expansion (\ref{Neumy}) with $\zeta=0.2$. Other parameters are $\alpha=v=1$.}
\label{fig3}
\end{figure}

Note that we recover the corresponding FPT densities for a totally absorbing wall \cite{Malakar18} by taking the limit $\kappa_0\rightarrow v$, which is equivalent to taking $\zeta\rightarrow  0$:
\numparts
\begin{eqnarray}
\label{f10}
\fl &\lim_{\kappa_0\rightarrow v}f_1(y,t) \\
\fl &=\frac{  \e^{-\alpha t}}{t+y/v}\bigg \{ \frac{\alpha y}{v}I_0(\alpha\sqrt{ t^2-y^2/v^2}) +\frac{\sqrt{t-y/v}}{\sqrt{t+y/v}}I_1(\alpha \sqrt{t^2-y^2/v^2})\bigg ]\Theta(\tau-y/v) ,\nonumber \\
\fl  &\lim_{\kappa_0\rightarrow v} f_{-1}(y,t)= \e^{-\alpha t}\bigg [  \frac{\alpha y/v}{\sqrt{t^2-y^2/v^2}}I_1(\alpha\sqrt{ t^2-y^2/v^2})\Theta(t-y/v)+ \delta(t-y/v)\bigg ].\nonumber \\
\fl 
\label{fm10}
\end{eqnarray}
\endnumparts
In Fig. \ref{fig3} we show sample plots of $f_{\pm 1}(x_0,t)$ for a fixed initial position $x_0$. We compare the results for a totally absorbing wall ($\zeta=0$) with a partially absorbing wall ($\zeta=0.2)$. In the latter case we use the first two terms in the Neumann series expansion (\ref{Neumy}). It can be seen that the FPT densities become independent of $\kappa_0$ in the large time limit, which is consistent with the large time asymptotics for $D>0$. As expected, a partially absorbing wall tends to enhance $f_1$ at the expense of $f_{-1}$ compared with a totally absorbing wall. That is, a particle that starts in the left-moving state and does not tumble is absorbed with unit probability in the case of a totally absorbing wall but can be reflected in the case of a partially absorbing wall.

 \section{Conclusion} In this paper we used stochastic calculus to derive the forward Kolmogorov equation for a diffusive  RTP on the half-line with a partially absorbing wall at $x=0$. We obtained two distinct boundary conditions whose zero-diffusion limits recovered the standard non-sticky and sticky boundary conditions, respectively. 
We assumed that the particle is absorbed when the amount of boundary-particle contact time (discrete or continuous local time) exceeded a quenched random threshold. We also derived the corresponding backward Kolmogorov equation and analysed the FPT densities in the case of a non-sticky wall. We end with a few comments.

\begin{enumerate}

\item As far as we are aware the boundary conditions (\ref{BCmodelA}) and (\ref{BCmodelB}) for a partially absorbing wall have not been considered previously. In the latter case (model B), setting $\kappa_0=0$ ($\kappa_0=\infty$) recovers the totally reflecting (totally absorbing) boundary conditions introduced in Ref \cite{Malakar18}. Here we analysed in more detail how one obtains the sticky boundary condition of a non-diffusive RTP in the singular limit $D\rightarrow 0$. The less intuitive boundary conditions (\ref{BCmodelA}) of model A are non-singular in the zero-diffusion limit, reducing to the corresponding non-sticky boundary condition of a non-diffusive RTP. 

\item Another major difference between models A and B is that the boundary-particle contact time is given by a discrete local time and a continuous local time, respectively. Hence, implementing a Robin-like boundary condition means taking the quenched local time threshold for absorption to be generated from a discrete geometric distribution in the case of model A and a continuous exponential distribution in the case of model B. In future work it would be interesting to explore more general choices of threshold distributions along analogous lines to previous encounter-based models of non-diffusive RTPs \cite{Bressloff22b,Bressloff23}.

\item It is important to distinguish between the sticky boundary condition for a non-diffusive RTP and for a pure Brownian particle. The notion of sticky reflecting Brownian motion (BM) originally arose from the construction of a general class of boundary conditions for the diffusion equation in $[0,\infty)$ that are consistent with stochastic processes behaving like standard BM in $(0,\infty)$ \cite{Feller52}. It was subsequently shown that sample paths of sticky reflecting BM can be generated using a random time change that slows down reflected paths so that the total time spent at the origin $x=0$ has positive Lebesgue measure \cite{Ito63}. A more recent formulation of sticky BM is based on the inclusion of a strongly localised attractive potential close to the boundary at $x=0$ \cite{Rabee20}. We subsequently developed an encounter-based version of the latter in order to model diffusion-mediated absorption at a sticky boundary \cite{Bressloff23a}. As part of our analysis we showed that there is no direct relationship between a sticky non-diffusive RTP and sticky BM. Within the context of the current paper, this means that the boundary conditions (\ref{BCmodelB}) of model B treat the wall as non-sticky for $D>0$ (beyond the effects of the drift term). A natural generalisation of the current paper would be to include some form of enhanced stickiness when $D>0$.

\end{enumerate}



\section*{References}
 
\begin{thebibliography}{9}



\bibitem{Angelani15} Angelani L 2015 Run-and-tumble particles, telegrapher's equation
and absorption problems with partially reflecting boundaries
{\em J. Phys. A: Math. Theor.} {\bf 48}, 495003 

\bibitem{Angelani17} Angelani L 2017 Confined run-and-tumble swimmers in one dimension. {\em J. Phys. A} {\bf 50}, 325601  

\bibitem{Angelani23} Angelani L 2023  One-dimensional run-and-tumble motions
with generic boundary conditions. {\em J. Phys. A} {\bf 56} ,455003.

 \bibitem{Rabee20} Bou-Rabee N and Holmes-Cerfom M C 2020 Sticky Brownian motion and its numerical solution. {\em SIAM Review} {\bf 62} 164-195

\bibitem{Bressloff22} Bressloff P C 2022  Diffusion-mediated absorption by partially reactive targets: Brownian functionals and generalized propagators. {\em J. Phys. A.} {\bf 55} 205001

 \bibitem{Bressloff22a} Bressloff P C 2022 Spectral theory of diffusion in partially absorbing media. {\em Proc. R. Soc. A} {\bf 478} 20220319
 
 \bibitem{Bressloff22b} Bressloff P C 2022 Encounter-based model of a run-and-tumble particle. {\em J. Stat. Mech.} {\bf 113206}  


\bibitem{Bressloff23} Bressloff P C 2023 Encounter-based model of a run-and-tumble particle II: absorption at sticky boundaries. {\em J. Stat. Mech.} {\bf 043208} 

\bibitem{Bressloff23a} Bressloff P C 2023 Close encounters of the sticky kind: Brownian motion at absorbing boundaries. {\em Phys. Rev. E} {\bf 107} 064121 

\bibitem{Bressloff25a} Bressloff P C 2025 Encounter-based model of a run-and-tumble particle with stochastic resetting {\em J. Phys. A} {\bf 58} 125002 (30pp)



\bibitem{Bressloff25b} Bressloff P C 2025 Stochastic calculus of run-and-tumble motion: an applied perspective. {\em Proc Roy Soc. A} {\bf 481} 20240815.


\bibitem{Dhar19} Dhar A, Kundu A, Majumdar S N, Sabhapandit S and 
Schehr G 2019 Run-and-tumble particle in one-dimensional confining
potentials: Steady-state, relaxation, and first-passage properties,
{\em Phys. Rev. E} {\bf 99}, 032132


  \bibitem{Evans18} Evans M R and Majumdar S N 2018 Run and tumble particle under resetting: a renewal approach. {\em J. Phys. A: Math. Theor.} {\bf 51} 475003 (2018).
  
\bibitem{Feller52} Feller W 1952 The parabolic differential equations and the associated semi-groups of transformations. {\em Ann. Math.} 468-519 
 
 
\bibitem{Grebenkov06} Grebenkov D S 2006 {Partially reflected Brownian motion: A stochastic
approach to transport phenomena.} in {\em Focus on Probability Theory}
Ed. Velle, L R pp. 135-169 (Hauppauge: Nova Science Publishers)

\bibitem{Grebenkov20} Grebenkov D S 2020  {Paradigm shift in diffusion-mediated surface phenomena.} {\em Phys. Rev. Lett.} {\bf 125} 078102  

\bibitem{Grebenkov22} Grebenkov D S 2022  {An encounter-based approach for restricted diffusion with a gradient drift.}  {\em J. Phys. A.} {\bf 55} 045203 

\bibitem{Grebenkov24} Grebenkov D S 2024 Encounter-based approach to target search problems
In {\em Target Search Problems}  Springer Nature Switzerland pp. 77-105 

\bibitem{Gueneau24} Gu\'eneau M and Touzo L 2024  Relating absorbing and hard wall boundary
conditions for a one-dimensional run-and-tumble
particle. {\em J. Phys. A: Math. Theor.} {\bf 57} 225005  

\bibitem{Gueneau25} Gu\'eneau M, Majumdar S N and Scher G 2025 Run-and-tumble particle in one-dimensional potentials: Mean first-passage time and applications. Phys. Rev. E {\bf 111} 014144

 \bibitem{Ito63} Ito K and McKean H P 1963 Brownian motions on a half line. Illinois {\em J. Math.} {\bf 7} 181-231

 \bibitem{Malakar18} Malakar K, Jemseena V, Kundu A, Vijay Kumar, 
Sabhapandit S, Majumdar S N, Redner S and Dhar A 2018 Steady state, relaxation and first-passage properties of a run-and-tumble particle in one-dimension, {\em J. Stat.
Mech.} {\bf 043215} 

 \bibitem{McKean75} McKean H P 1975 {Brownian local time.} {\em Adv. Math.} {\bf 15} 91-111  
 
 


 \bibitem{Mori20} Mori F, Le Doussal P, Majumdar S N and Schehr G. 2020 Universal Survival Probability for a d-Dimensional Run-and-Tumble Particle. {\em Phys. Rev. Lett.} {\bf 124}, 090603
 
 

\bibitem{Pal23} Mallikarjun R and Pal A. 2023 Chiral run-and-tumble walker: Transport and optimizing
search {\em Physica A} {\bf 622}, 128821

  
\bibitem{Santra20} Santra I, Basu U and Sabhapandit S. 2020 Run-and-tumble particles in two-dimensions: Marginal position distributions {\em Phys. Rev. E} {\bf 101} 062120. 
 
\bibitem{Santra20a} Santra I, Basu U and Sabhapandit S 2020 Run-and-tumble particles in two dimensions under stochastic resetting conditions {\em J. Stat. Mech.} 113206
 
 \bibitem{Singh20}  Singh P, Sabhapandit S and Kundu A 2020 Run-and-tumble particle in inhomogeneous media in one dimension  {\em J. Stat. Mech.} {\bf 083207} 
 
 \bibitem{Doussal22} Smith N R, Le Doussal P, Majumdar S N and Schehr G 2022 Exact position distribution of a harmonically confined run-and-tumble particle in two dimensions {\em Phys. Rev. E} {\bf 106} 054133


 

 



 


 

 




\end{thebibliography}
\end{document}